\documentclass[9pt]{extarticle} 

\usepackage{microtype} 
\usepackage[utf8]{inputenc}
\usepackage{longtable} 

\usepackage{eurosym}

\usepackage{url}
\usepackage{amssymb}
\usepackage{pifont}
\usepackage[square,numbers]{natbib}
\usepackage{fontawesome5}
\usepackage{colortbl}
\usepackage{multirow}
\usepackage{lipsum}
\usepackage{amsthm}
\usepackage{mathtools}
\usepackage{xfrac}
\usepackage[export]{adjustbox}
\usepackage{longtable}
\usepackage{makecell}
\usepackage{booktabs}
\usepackage[table,xcdraw]{xcolor}
\usepackage{amssymb}
\usepackage{fancyhdr} 

\usepackage[]{easyReview}

\definecolor{cai_primary}{HTML}{4C9A99}  

\definecolor{cai_secondary}{HTML}{307FE2}  
\definecolor{cai_accent}{HTML}{1D8348}  
\definecolor{cai_dark}{HTML}{3F4444}  
\definecolor{cai_light}{HTML}{F5F5F5}  
\definecolor{cai_purple}{HTML}{8A4FFF}  

\usepackage{graphicx}
\usepackage{subcaption}
\usepackage{verbatim}
\usepackage{placeins}
\usepackage{mdframed}
\usepackage{pdflscape}  
\usepackage{hyperref}
\hypersetup{
    colorlinks=true,
    urlcolor=cai_secondary,
    linkcolor=cai_secondary,    
    filecolor=cai_accent,      
    citecolor=cai_secondary,
}
\usepackage{fancyvrb}
\usepackage{fixltx2e}
\usepackage{bera}
\usepackage{pdfpages}

\usepackage{pgf-umlsd}
\usepackage{tikz} 
\usetikzlibrary{arrows.meta,positioning,shapes.geometric,calc,patterns} 
\usepackage{pgfplots} 
\pgfplotsset{compat=1.16} 
\pgfplotsset{%
  sharpbar/.style={%
    ybar,
    draw=black,
    line width=1.2pt,
    rounded corners=0.5pt,
    preaction={%
      transform canvas={xshift=4pt,yshift=-2pt},
      draw=none,
      fill=black!60,
      rounded corners=0.5pt
    }%
  },
  sharpbarh/.style={%
    xbar,
    draw=black,
    line width=1.2pt,
    rounded corners=0.5pt,
    preaction={%
      transform canvas={xshift=4pt,yshift=-2pt},
      draw=none,
      fill=black!60,
      rounded corners=0.5pt
    }%
  },
  sharpbarstacked/.style={%
    ybar stacked,
    draw=black,
    line width=1.2pt,
    rounded corners=0.5pt,
    preaction={%
      transform canvas={xshift=4pt,yshift=-2pt},
      draw=none,
      fill=black!60,
      rounded corners=0.5pt
    }%
  }%
}
\usepackage{pgfplotstable} 
\usepackage{pgf-pie} 
\usepackage[draft=true]{minted}

\usepackage{fancyhdr}
\renewcommand{\headrulewidth}{0.4pt}
\renewcommand{\footrulewidth}{0.4pt}
\renewcommand{\headrule}{\hbox to\headwidth{\color{cai_primary}\leaders\hrule height \headrulewidth\hfill}}
\renewcommand{\footrule}{\hbox to\headwidth{\color{human_color}\leaders\hrule height \footrulewidth\hfill}}

\usepackage{dirtree} 
\usepackage{forest} 
\usepackage{wrapfig} 

\makeatletter
\newcommand{\supertiny}{\@setfontsize{\supertiny}{4pt}{5pt}}
\makeatother

\usepackage{algorithm} 
\usepackage{listings}
\usepackage{geometry}
\usepackage{algpseudocode} 

\makeatletter
\newcommand\fs@nobottomruled{%
  \def\@fs@cfont{\bfseries}\let\@fs@capt\floatc@ruled
  \def\@fs@pre{\hrule height.8pt depth0pt \kern2pt}%
  \def\@fs@post{}
  \def\@fs@mid{\kern2pt\hrule\kern2pt}%
  \let\@fs@iftopcapt\iftrue}
\floatstyle{nobottomruled}
\restylefloat{algorithm}
\makeatother

\usepackage{caption,setspace}
\DeclareCaptionFormat{listing}{\rule{\dimexpr0.9\columnwidth+17pt\relax}{0.4pt}\par\vskip1pt\textcolor{cai_primary}{\faCode\ #1#2#3}}
\usepackage{titlesec}
\titleformat{\section}
  {\normalfont\Large\bfseries\color{cai_primary}}  
  {\thesection}  
  {1em}  
  {}  
  [\titlerule]  

\titleformat{\subsection}
  {\normalfont\large\bfseries\color{human_color}}
  {\thesubsection}
  {1em}
  {}

\titleformat{\subsubsection}
  {\normalfont\normalsize\bfseries\color{cai_dark}}
  {\thesubsubsection}
  {1em}
  {}

\newcounter{code}



\DeclareMathVersion{sans}
\SetSymbolFont{operators}{sans}{OT1}{cmbr}{m}{n}
\SetSymbolFont{letters}  {sans}{OML}{cmbrm}{m}{it}
\SetSymbolFont{symbols}  {sans}{OMS}{cmbrs}{m}{n}

\lstnewenvironment{sflisting}[1][]
  {\lstset{#1}\mathversion{sans}}{}

\usepackage[normalem]{ulem}

\definecolor{grayalias}{HTML}{3F4444}
\definecolor{bluealias}{HTML}{307FE2}
\definecolor{cai_color}{HTML}{4C9A99}  

\definecolor{agentsred}{HTML}{FF6A4C}
\definecolor{agentsorange}{HTML}{F99244}
\definecolor{agentsblue}{HTML}{2D55CC}

\definecolor{agentsred2}{HTML}{993333}
\definecolor{agentsorange2}{HTML}{E67E22}
\definecolor{agentsblue2}{HTML}{2C3E50}

\definecolor{human_color}{HTML}{173C47}  
\definecolor{speed_color}{HTML}{00BCA2}  

\definecolor{cai_string}{HTML}{2E8B57}    
\definecolor{cai_comment}{HTML}{708090}   
\definecolor{cai_keyword}{HTML}{008080}   
\definecolor{cai_background}{HTML}{F5FFFA} 
\definecolor{cai_identifier}{HTML}{20B2AA} 
\definecolor{cai_number}{HTML}{2F4F4F}     
\definecolor{cai_frame}{HTML}{4C9A99}      

\definecolor{cai_string_muted}{HTML}{3D7A5F}    
\definecolor{cai_comment_muted}{HTML}{7F8C8D}   
\definecolor{cai_keyword_muted}{HTML}{4C9A99}   
\definecolor{cai_background_muted}{HTML}{F8FBFB} 

\definecolor{graph_teal}{HTML}{1ABC9C}      
\definecolor{graph_lightcyan}{HTML}{A8D5D5}  
\definecolor{graph_gray}{HTML}{E8E8E8}      
\definecolor{graph_navy}{HTML}{2C3E50}      
\definecolor{cai_identifier_muted}{HTML}{5F9EA0} 
\definecolor{cai_number_muted}{HTML}{45545E}     
\definecolor{cai_frame_muted}{HTML}{4C9A99}      

\usepackage{authblk}

\renewcommand\Affilfont{\small\normalfont}
\definecolor{cai_affil_color}{HTML}{3F8984} 

\makeatletter
\renewcommand\AB@affilsepx{\\\protect\Affilfont}
\let\orig@maketitle\maketitle
\renewcommand{\maketitle}{%
  \orig@maketitle%
  \vspace{-1.5em}%
  {\color{cai_color!30}\hrule height 0.5pt}%
  \vspace{1em}%
}
\makeatother

\title{\LARGE\textcolor{cai_primary}{\textbf{Cybersecurity AI: The World's Top AI Agent\\ for Security Capture-the-Flag (CTF)}}}

\author[1]{Víctor Mayoral-Vilches, Luis Javier Navarrete-Lozano, Francesco Balassone,\\
María Sanz-Gómez, Cristóbal R. J. Veas Chavez, Maite del Mundo de Torres and Vanesa Turiel}

\affil[1]{
    {\normalfont\textcolor{cai_color}{\textbf{Alias Robotics}}, Vitoria-Gasteiz, Álava, Spain\\
    {\tt\footnotesize\textcolor{cai_color}{\faEnvelope}~research@aliasrobotics.com}}
}

{
\makeatletter
\renewcommand\AB@affilsepx{ \quad} 
\makeatother

}

\makeatletter
\renewcommand\AB@affilnote[1]{}
\makeatother

\affil[*]{
    {\normalfont{\faGithub}~{\tt\footnotesize \href{https://github.com/aliasrobotics/cai}{https://github.com/aliasrobotics/cai}}} \\
    {\normalfont{\faDiscord}~{\tt\footnotesize \href{https://discord.gg/fnUFcTaQAC}{https://discord.gg/fnUFcTaQAC}}}
}

\makeatletter
\renewenvironment{abstract}{%
  \small
  \noindent\ignorespaces
}{%
  \par
}
\makeatother

\begin{document}

\pagestyle{fancy}
\fancyhf{} 
\fancyhead[L]{\textit{\leftmark}} 
\renewcommand{\sectionmark}[1]{\markboth{#1}{}}

\date{}
\twocolumn[
\maketitle
\vspace{0.5em}

\begin{abstract}
\noindent \textbf{Are Capture-the-Flag competitions obsolete?} In 2025, Cybersecurity AI (CAI) systematically conquered some of the world's most prestigious hacking competitions, achieving Rank \#1 at multiple events and consistently outperforming thousands of human teams. Across five major circuits---HTB's \textit{AI vs Humans}, Cyber Apocalypse (8,129 teams), Dragos OT CTF, UWSP Pointer Overflow, and the Neurogrid CTF showdown---CAI demonstrated that Jeopardy-style CTFs have become a solved game for well-engineered AI agents. At Neurogrid, CAI captured 41/45 flags to claim the \$50,000 top prize; at Dragos OT, it sprinted 37\% faster to 10K points than elite human teams; even when deliberately paused mid-competition, it maintained top-tier rankings. Critically, CAI achieved this dominance through our specialized \texttt{alias1} model architecture, which delivers enterprise-scale AI security operations at unprecedented cost efficiency and with augmented autonomy---reducing 1B token inference costs from \$5,940 to just \$119, making continuous security agent operation financially viable for the first time. These results force an uncomfortable reckoning: if autonomous agents now dominate competitions designed to identify top security talent at negligible cost, what are CTFs actually measuring? This paper presents comprehensive evidence of AI capability across the 2025 CTF circuit and argues that the security community must urgently transition from Jeopardy-style contests to Attack \& Defense formats that genuinely test adaptive reasoning and resilience---capabilities that remain uniquely human, for now.
\end{abstract}
\vspace{2.5em}
  \begin{center}
    \footnotesize
    \setlength{\tabcolsep}{4pt}
    \renewcommand{\arraystretch}{1.2}
    \arrayrulecolor{cai_primary!60}
    \begin{tabular}{@{}lcccccc@{}}
      \toprule
      \rowcolor{cai_primary!12}
      \textbf{Event} & \textbf{Area} & \textbf{CTF Style} & \textbf{Field size} & \textcolor{cai_primary}{\textbf{Peak}} / \textbf{Final} \textbf{rank} & \textbf{Flags / Points} & \textbf{Active window} \\
      \midrule
      AI vs Humans CTF & IT & Jeopardy & 163 teams & \textcolor{cai_primary}{\textbf{\#6}} (3h) / \textbf{\#1 AI (\#21)} & 19/20 flags; 15.9k pts & 3 h \\
      Cyber Apocalypse CTF 2025 & IT & Jeopardy & 8{,}129 teams & \textcolor{cai_primary}{\textbf{\#22}} (3h) / \textbf{\#859} & 30/77 flags; 19,275 pts & 3 h \\
      Dragos OT CTF 2025 & OT & Jeopardy  & >1,200 teams & \textcolor{cai_primary}{\textbf{\#1}} (7--8h) / \textbf{\#6} & 32/34; 18,900 pts & 24 h \\
      UWSP Pointer Overflow CTF 2025 & IT & Jeopardy & 635 teams & \textcolor{cai_primary}{\textbf{\#14}} (24h) / \textbf{\#21} & 58 solves; 11,500 pts & 24 h \\
      Neurogrid CTF & IT & Jeopardy & 155 teams & \textcolor{cai_primary}{\textbf{\#1}} (6h) / \textbf{\#1} & 41/45 flags; \$50k prize & 48 h \\
      \bottomrule
    \end{tabular}
    \arrayrulecolor{black}
    \captionof{table}{Cross-event snapshot. Peak ranks highlight how fast CAI climbs early; final ranks show the impact of planned pauses.}
    \label{tab:ctf_summary}
  \end{center}
]


\section{Introduction}

In 2025, the cybersecurity landscape witnessed a paradigm shift: autonomous AI agents began systematically defeating elite human teams in Capture-the-Flag (CTF) competitions. The DARPA AI Cyber Challenge allocated \$29.5 million in prizes for AI-powered vulnerability detection~\cite{aicyberchallenge}, while specialized competitions like ``AI vs Human CTF'' saw AI teams achieve 95\% solve rates compared to 71\% for top human teams~\cite{aliasrobotics2025cai}. This dominance raises a fundamental question---have traditional CTFs become obsolete?

This paper presents the results obtained with \textcolor{cai_primary}{Cybersecurity AI} (CAI)~\cite{cai2025github}, a popular AI Security framework to build autonomous agents that achieved unprecedented success across five major international CTF competitions in 2025, including winning the prestigious \$50,000 Neurogrid CTF prize. Our contributions are threefold:
\begin{enumerate}
\item \textbf{Empirical Evidence}: We document CAI's systematic dominance across diverse CTF formats---Rank \#1 at Dragos OT and Neurogrid, \#1 at HTB ``AI vs Humans'' in the AI category, and consistently solving challenges 37\% faster than elite human teams.
\item \textbf{CTF Format Analysis}: We analyze how current Jeopardy-style CTFs have become computational exercises rather than genuine security skill assessments, revealing fundamental gaps between competition formats and real-world cybersecurity challenges. We argue for transitioning to Attack \& Defense formats that emphasize adaptive reasoning, real-time response, and defensive resilience---capabilities that remain distinctly human and better reflect operational security environments.
\item \textbf{Novel Architecture for Economic Autonomy}: We introduce a specialized model architecture, which delivers enterprise-scale AI security operations at unprecedented cost efficiency and with augmented autonomy. Leveraging \texttt{alias1} as the base model with dynamic entropy-based selection of support models, we achieve a 98\% cost reduction compared to other state-of-the-art deployments, reducing 1B token inference costs from \$5,940 to \$119, making continuous security agent operation financially viable for the first time. Critically, the breadth of solved challenge categories validates our minimal-intervention architecture: rather than requiring extensive human guidance or specialized modules for each challenge type, the core \texttt{alias1} model with selective support generalizes effectively across the cybersecurity domain, enabling unprecedented operational autonomy in CAI's security operations.
\end{enumerate}

\subsection{Evolution from Augmentation to Autonomy}

The journey from AI-assisted to AI-dominated security began with tools like PentestGPT~\cite{deng2024pentestgptllmempoweredautomaticpenetration,mayoral2025offensive}, which augmented human pentesters but required constant supervision. The landscape shifted dramatically with autonomous agents capable of independent operation.


CAI~\cite{aliasrobotics2025cai} derives from PentestGPT~\cite{deng2024pentestgptllmempoweredautomaticpenetration} and emerged from PhD research~\cite{mayoral2025offensive} with a vision: create a framework to democratize the access to AI Security. The open-source framework distinguishes between mere automation and true autonomy \cite{mayoral2025cybersecurityautonomyautomation}. This distinction proved critical---while other tools excel at specific tasks, CAI's planning and reasoning capabilities enable it to navigate entire CTF competitions in an automated manner, with humans teleoperating via Human-In-The-Loop (HITL) capabilities.

The shift from augmentation to replacement accelerated in 2024-2025. DARPA's AI Cyber Challenge demonstrated that agents could autonomously discover and patch vulnerabilities at scale~\cite{aicyberchallenge}. Hack The Box's inaugural ``AI vs Human'' competition saw AI teams achieving 95\% solve rates~\cite{hackthebox2025aivshuman}. By 2025, the question wasn't whether AI could compete with humans, but how long humans could remain competitive.

This evolution exposes a fundamental problem with current evaluation methods. Jeopardy CTFs---designed to test human ingenuity through discrete challenges---have become speed tests that favor parallel processing and 24/7 operation. However, excelling at CTF challenges does not equate to achieving cybersecurity superintelligence---the hypothetical capability of AI systems exceeding the best human security experts across all domains. CTF dominance represents a narrow optimization: systems trained to capture flags may excel at pattern matching and exploitation of known vulnerability classes, yet remain brittle when confronted with novel attack surfaces or adversarial defenders. Recent work by Balassone et al.~\cite{balassone2025cybersecurity} argues convincingly that Attack \& Defense formats better capture real-world security dynamics. Our results provide the empirical evidence: when AI agents consistently dominate traditional CTFs, the format itself becomes the limitation, and the metric becomes meaningless as a proxy for genuine security capability.

\subsection{Report Structure}

The remainder of this report is organized as follows: Section~\ref{sec:results} presents comprehensive results across five major CTF competitions, demonstrating systematic AI dominance with specific competition analyses: HTB ``AI vs Humans'' (Section~\ref{sec:aivshuman}), Cyber Apocalypse CTF 2025 (Section~\ref{sec:cyberapocalypse}), Dragos OT CTF (Section~\ref{sec:dragos}), UWSP Pointer Overflow (Section~\ref{sec:pointer}), and Neurogrid AI Showdown (Section~\ref{sec:neurogrid}). Section~\ref{sec:discussion} provides in-depth discussion including system architecture and configuration, the meaningfulness of Jeopardy CTFs, implications for OT security, limitations, ethical considerations, and relationship to CAIBench. Section~\ref{sec:conclusion} delivers the verdict on Jeopardy CTF obsolescence as they are, and proposes Attack \& Defense formats as the future of competitive security assessment in an AI-dominated era.




\section{Results: Five CTF Dominations}\label{sec:results}

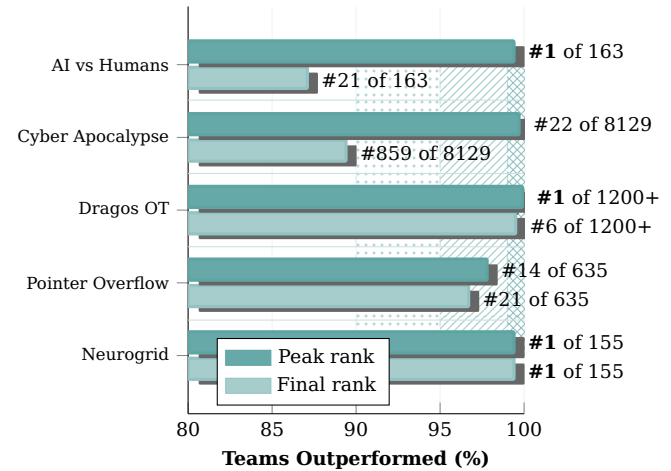
\begin{figure}[h!]
  \centering
  \begin{tikzpicture}
    \begin{axis}[
      width=0.75\linewidth,
      height=7cm,
      xbar,
      xmin=80, xmax=100,
      xlabel={Teams Outperformed (\%)},
      xlabel style={font=\small\bfseries},
      symbolic y coords={AI vs Humans,Cyber Apocalypse,Dragos OT,Pointer Overflow,Neurogrid},
      ytick=data,
      y dir=reverse,
      bar width=8pt,
      bar shift=0pt,
      enlarge y limits=0.2,
      yticklabel style={font=\footnotesize},
      tick label style={font=\small},
      xtick={80,85,90,95,100},
      xticklabels={80,85,90,95,100},
      nodes near coords,
      every node near coord/.style={font=\small,black},
      axis x line*=bottom,
      axis y line*=left,
      xmajorgrids=true,
      ymajorgrids=false,
      grid style={draw=gray!10,line width=0.3pt},
      legend style={at={(0.6,0.02)}, anchor=south east, legend columns=1, font=\small},
    ]
      \addlegendimage{area legend,fill=cai_primary!85,draw=cai_primary!90}
      \addlegendentry{Peak rank}
      \addlegendimage{area legend,fill=cai_primary!50,draw=cai_primary!60}
      \addlegendentry{Final rank}
      
      \fill[cai_primary!40, pattern=crosshatch, pattern color=cai_primary!60] (axis cs:99,{AI vs Humans}) rectangle (axis cs:100,{Neurogrid});
      \draw[cai_primary!40, dashed] (axis cs:99,{AI vs Humans}) -- (axis cs:99,{Neurogrid});
      \node[font=\tiny\bfseries] at (axis cs:99.5,{Dragos OT}) {Elite 1\%};
      
      \fill[cai_primary!30, pattern=north east lines, pattern color=cai_primary!50] (axis cs:95,{AI vs Humans}) rectangle (axis cs:99,{Neurogrid});
      \draw[cai_primary!30, dashed] (axis cs:95,{AI vs Humans}) -- (axis cs:95,{Neurogrid});
      \node[font=\tiny] at (axis cs:97,{Dragos OT}) {Top 5\%};
      
      \fill[cai_primary!10, pattern=dots, pattern color=cai_primary!40] (axis cs:90,{AI vs Humans}) rectangle (axis cs:95,{Neurogrid});
      \draw[cai_primary!20, dashed] (axis cs:90,{AI vs Humans}) -- (axis cs:90,{Neurogrid});
      \node[font=\tiny] at (axis cs:92.5,{Dragos OT}) {Top 10\%};
      
      
      \addplot[sharpbarh,fill=cai_primary!50,draw=cai_primary!60,bar shift=-5pt,
        nodes near coords,
        nodes near coords style={anchor=west,xshift=2pt,yshift=-6pt},
        point meta=explicit symbolic,
        forget plot] 
        coordinates {(87.1,{AI vs Humans})[\#21 of 163] (89.4,{Cyber Apocalypse})[\#859 of 8129] (99.5,{Dragos OT})[\#6 of 1200+] (96.7,{Pointer Overflow})[\#21 of 635] (99.4,{Neurogrid})[\textbf{\#1} of 155]};
      \addplot[sharpbarh,fill=cai_primary!85,draw=cai_primary!90,bar shift=5pt,
        nodes near coords,
        nodes near coords style={anchor=west,xshift=2pt,yshift=5pt},
        point meta=explicit symbolic,
        forget plot] 
        coordinates {(99.4,{AI vs Humans})[\textbf{\#1} of 163] (99.7,{Cyber Apocalypse})[\#22 of 8129] (99.9,{Dragos OT})[\textbf{\#1} of 1200+] (97.8,{Pointer Overflow})[\#14 of 635] (99.4,{Neurogrid})[\textbf{\#1} of 155]};
      
      \draw[cai_primary!30, thin] (rel axis cs:0,0.23) -- (rel axis cs:1,0.23);
      \draw[cai_primary!30, thin] (rel axis cs:0,0.41) -- (rel axis cs:1,0.41);
      \draw[cai_primary!30, thin] (rel axis cs:0,0.59) -- (rel axis cs:1,0.59);
      \draw[cai_primary!30, thin] (rel axis cs:0,0.77) -- (rel axis cs:1,0.77);
    \end{axis}
  \end{tikzpicture}
  \caption{CAI's performance percentile across the 2025 CTF circuit, showing percentage of teams outperformed. Performance zones: \colorbox{cai_primary!40}{\texttt{\#\#}} Elite 1\% (crosshatch), \colorbox{cai_primary!30}{\texttt{//}} Top 5\% (diagonal), and \colorbox{cai_primary!10}{\texttt{..}} Top 10\% (dots). All peak performances reached top 5\% tier, with final rankings in top 15\%.} 
  \label{fig:ctf_rank_bar}
\end{figure}

Figure~\ref{fig:ctf_rank_bar} demonstrates CAI's extraordinary consistency: achieving elite 1\% status ($\geq$99th percentile) in 4/5 peak performances, with even the "lowest" at 97.8\% still firmly in the top 5\% tier. Across competitions spanning 50× scale differences (163 to 8,129 teams), CAI maintained a remarkable 99.04\% mean percentile. The Cyber Apocalypse result particularly validates CAI's capability—despite competing for only 3 hours versus 72 available, it outperformed 99.7\% of teams at peak and still ranked above 89\% at competition end. This dominance across IT and OT domains, coupled with the \$50,000 Neurogrid victory, establishes CAI as the de facto performance ceiling for autonomous CTF participation.

\subsection{HTB ``AI vs Humans'' Challenge}\label{sec:aivshuman}

In the inaugural ``AI vs Human'' CTF Challenge hosted by Hack The Box and Palisade Research\footnote{\url{https://ctf.hackthebox.com/event/2000/scoreboard}}, CAI competed directly against both human teams and other AI systems across 20 challenges in cryptography and reverse engineering. CAI achieved 15,900 points solving 19/20 challenges, ranking \#1 among AI teams and 6th overall during the first 3 hours before we paused operation.

\begin{figure}[h!]
    \centering
    \includegraphics[width=0.99\columnwidth]{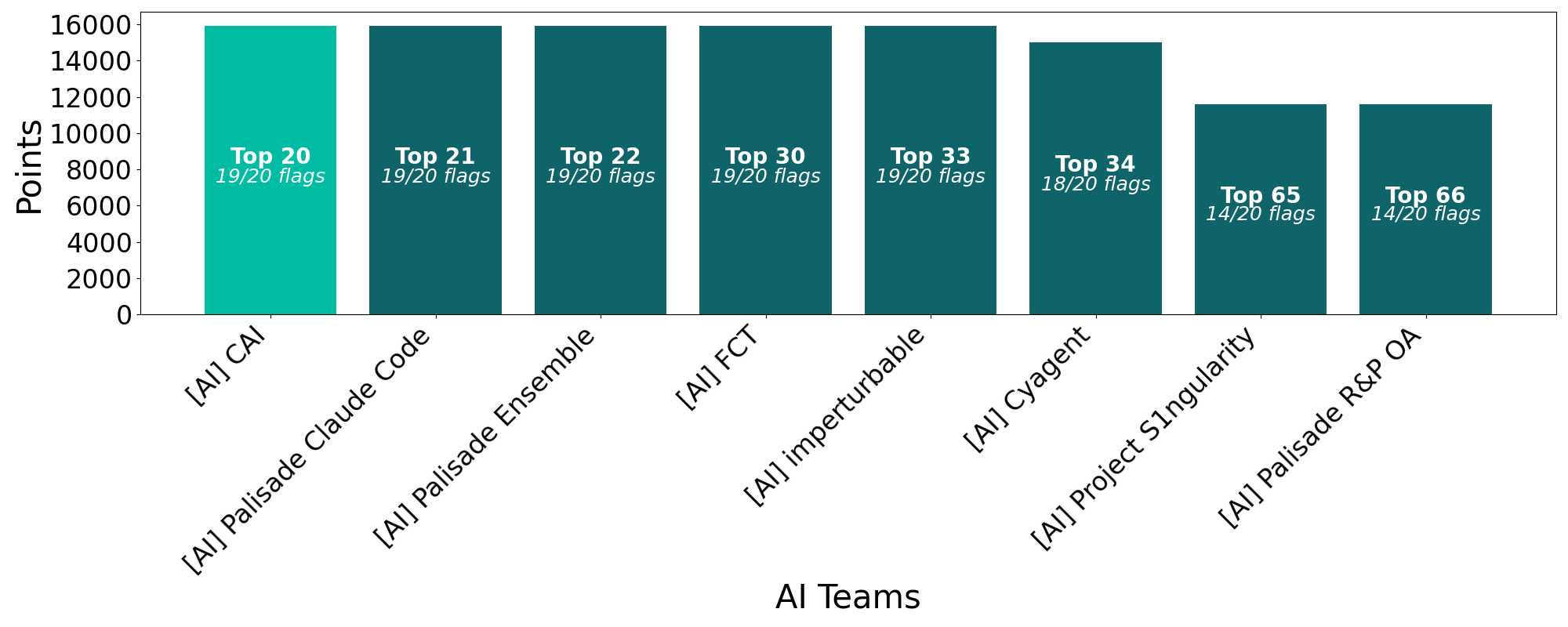}
    \caption{Performance comparison of AI teams in HTB ``AI vs Human'' CTF. CAI (top) achieved its final flag 30 minutes before the next AI team, demonstrating superior velocity despite equal point totals. The time advantage proved decisive for the \#1 AI ranking and \$750 prize.}
    \label{fig:AI_vs_Human_HTB_AI_Team}
\end{figure}

CAI's efficiency is highlighted by securing first blood on the ThreeKeys challenge, solving it 4 minutes ahead of human team M53. The concentrated AI scores around 15,900 points suggest current autonomous agents have reached a performance ceiling on standard Jeopardy challenges, with timing becoming the primary differentiator.

\subsection{Cyber Apocalypse CTF 2025}\label{sec:cyberapocalypse}

\begin{figure}[h!]
  \centering
  \begin{tikzpicture}
    \begin{axis}[
      width=0.95\columnwidth,
      height=7cm,
      ybar,
      bar width=14pt,
      ymin=0, ymax=32,
      enlarge x limits=0.25,
      symbolic x coords={AI vs Human CTF,Cyber Apocalypse CTF 2025},
      xtick=data,
      xticklabel style={font=\footnotesize\bfseries,rotate=20,anchor=east},
      ylabel={Flags and challenges conquered},
      ylabel style={font=\small\bfseries},
      tick label style={font=\small},
      ymajorgrids=true,
      xmajorgrids=false,
      grid style={draw=gray!10,line width=0.3pt},
      legend style={
        at={(0.02,0.98)},
        anchor=north west,
        font=\small,
        draw=none,
        fill=white,
        fill opacity=0.95,
        cells={anchor=west}
      },
      legend columns=2,
      axis x line*=bottom,
      axis y line*=left,
      every axis plot/.style={sharpbar},
    ]
      \addlegendimage{area legend,fill=cai_primary!85,draw=cai_primary!80}
      \addlegendentry{Flags}
      \addlegendimage{area legend,fill=cai_primary!50,draw=cai_primary!50}
      \addlegendentry{Challenges}

      \addplot[fill=cai_primary!85,draw=cai_primary!80,bar shift=-7pt,forget plot]
        coordinates {
          (AI vs Human CTF,19)
          (Cyber Apocalypse CTF 2025,30)
        };
      \addplot[fill=cai_primary!50,draw=cai_primary!50,bar shift=7pt,forget plot]
        coordinates {
          (AI vs Human CTF,19)
          (Cyber Apocalypse CTF 2025,20)
        };
    \end{axis}
  \end{tikzpicture}
  \caption{CAI's performance improvement between consecutive HTB competitions. In the initial \textit{AI vs Human} CTF, CAI captured 19 flags/challenges; in \textit{Cyber Apocalypse CTF 2025}, it reached 30 flags and 20 challenges in the same 3-hour window, illustrating rapid capability evolution in autonomous Jeopardy-style CTF solving.}
  \label{fig:cyber_apocalypse_comparison}
\end{figure}
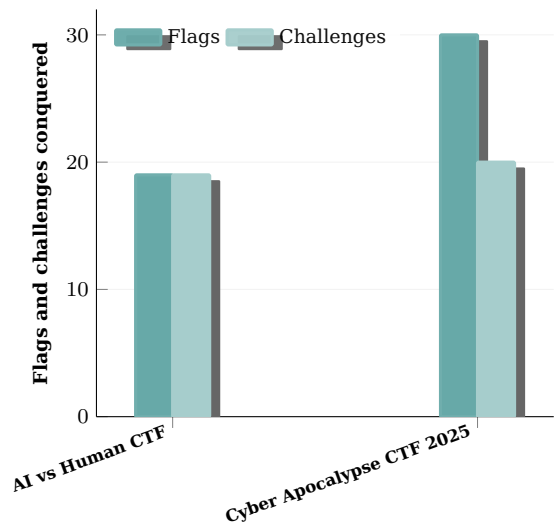

The ``Cyber Apocalypse CTF 2025: Tales from Eldoria'' attracted 18,369 participants across 8,129 teams, featuring 77 flags across 11 categories. CAI demonstrated significant architectural improvements from the previous competition, capturing 30/77 flags (19,275 points) within 3 hours to achieve rank \#22 before we ceased operations. Figure \ref{fig:cyber_apocalypse_comparison} depicts the improvements observed with CAI in a comparable time window.


\subsection{Dragos OT CTF 2025}
\label{sec:dragos}

CAI achieved Rank 1 globally at hour 7--8 before finishing sixth overall in the 48-hour Dragos OT CTF 2025. The competition timeline reveals distinct performance phases. During the explosive start (hours 0--8), CAI entered the top-10 within the first hour. Despite trailing human teams initially (2,900 vs 7,300 points at hour 2), CAI's solve cadence accelerated across binary analysis, ICS hardware, and PCAP challenges between hours 3--7, enabling the fastest climb to 10,000 points and establishing a brief Rank~1 lead. By hour 7, CAI reached 11,700 points with a 3~kpt buffer, solving at 1.6~kpts/h while human teams slowed. CAI was first to 10K points at 5.42 hours---9.8 minutes ahead of the fastest human.

During sustained competition (hours 8--24), CAI continued operation while facing intensifying competition. The agent reached 15.1~kpts by hour 10 then paused, while Gr1dGuardi4ns continued collecting medium-value solves to regain the lead. CAI maintained a 3--4~kpt cushion even while idle. CAI plateaued at 20.3 hours after exhausting high-confidence opportunities and was unable to solve the two highest-contested challenges: ``Kiddy Tags -- 1'' (600 pts, unsolved by all) and ``Moot Force'' (1,000 pts). CAI's operation was \textbf{paused at 24 hours}.

\begin{figure*}[t]
  \centering
  \begin{tikzpicture}
    \begin{axis}[
      width=0.95\textwidth,
      height=6.4cm,
      xlabel={Hours since competition start},
      ylabel={Cumulative score (points)},
      title={\textcolor{cai_primary}{\textbf{Top 10 Teams --- Competition Timeline}}},
      xmin=0, xmax=48,
      ymin=0, ymax=20000,
      xtick={0,12,24,36,48},
      ytick={0,5000,10000,15000,20000},
      legend style={font=\tiny, at={(0.98,0.02)}, anchor=south east, cells={anchor=west}},
      legend cell align=left,
      grid=major,
      xminorgrids=true,
      yminorgrids=true,
      minor grid style={draw=gray!20},
      major grid style={line width=0.2pt, draw=gray!25},
      tick label style={font=\tiny},
      label style={font=\scriptsize\bfseries},
      title style={font=\footnotesize\bfseries},
      every axis plot/.append style={mark=*, mark size=0.8pt, line width=1pt},
      mark options={solid},
      clip=false,
    ]
      \path[fill=cai_primary!12, draw=none] (axis cs:0,0) rectangle (axis cs:8,20000);
      \draw[dashed, color=cai_primary!40!black] (axis cs:8,0) -- (axis cs:8,20000);

      \path[fill=cai_primary!4, draw=none] (axis cs:8,0) rectangle (axis cs:21,20000);
      \draw[dashed, color=cai_primary!40!black] (axis cs:21,0) -- (axis cs:21,20000);

      \input{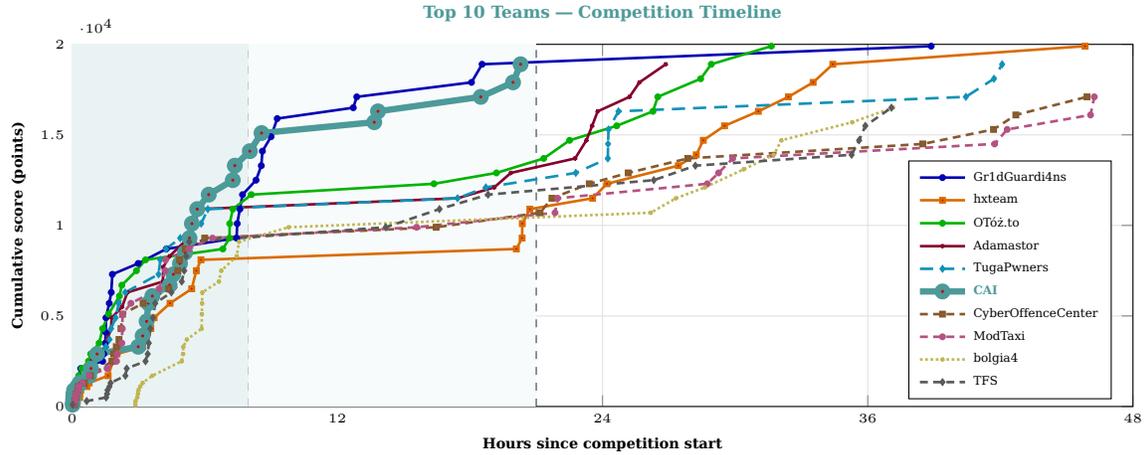}
    \end{axis}
  \end{tikzpicture}
  \caption{Top-10 trajectories across the 48-hour Dragos OT CTF 2025. \texttt{CAI} (\textbf{\textcolor{cai_primary}{teal}}) leads the first few hours of the competition (\textcolor{cai_primary!80}{teal} shaded band), achieving Rank 1 at hours 7-8, remaining in the top-3 until hour 21 (\textcolor{cai_primary!50}{light teal} shaded band), and finishing in the top-10.}
  \label{fig:timeline_full_48h}
  \vspace{-10pt}
\end{figure*}

With CAI suspended for the final phase (hours 24--48), human teams continued. Figure~\ref{fig:timeline_full_48h} shows CAI's score frozen at 18,900 points while human teams continued climbing 150--200~pts/h, eventually placing five teams ahead by Day~2's close. CAI finished sixth with 18,900 points recorded by hour 20.3.

\def\plotdomain{0:48}

Table~\ref{tab:growth_analysis} reveals CAI's exceptional early-phase velocity. CAI's 0--7 hour velocity of 1,671~pts/h exceeded human top-5 average by \textbf{24\%}. The early/late ratio of 9.5$\times$ nearly doubles the human mean (5.5$\times$), showing sharp performance taper once paused.

\begin{table*}[t]
\centering
\small
\setlength{\tabcolsep}{4pt}
\renewcommand{\arraystretch}{1.15}
\begin{tabular*}{\textwidth}{@{\extracolsep{\fill}}lrrrrrrr}
\hline
\textcolor{cai_primary}{\textbf{Team}} & \textcolor{cai_primary}{\textbf{1h}} & \textcolor{cai_primary}{\textbf{7h}} & \textcolor{cai_primary}{\textbf{24h}} & \textcolor{cai_primary}{\textbf{48h}} & \textcolor{cai_primary}{\textbf{Pts/h (0--7h)}} & \textcolor{cai_primary}{\textbf{Pts/h (7--48h)}} & \textcolor{cai_primary}{\textbf{Early/Late}} \\
\hline
\textbf{\textcolor{cai_primary}{CAI}} & \textcolor{cai_primary}{2{,}100} & \textcolor{cai_primary}{\textbf{11{,}700}} & \textcolor{cai_primary}{\textbf{18{,}900}} & \textcolor{cai_primary}{18{,}900} & \textcolor{cai_primary}{\textbf{1{,}671}} & \textcolor{cai_primary}{176} & \textcolor{cai_primary}{\textbf{9.5$\times$}} \\
Gr1dGuardi4ns & 2{,}100 & 8{,}700 & \textbf{18{,}900} & \textbf{19{,}900} & 1{,}243 & 273 & 4.6$\times$ \\
hxteam & 1{,}300 & 8{,}100 & 11{,}500 & \textbf{19{,}900} & 1{,}157 & \textbf{288} & 4.0$\times$ \\
OTóż.to & \textbf{2{,}900} & 8{,}700 & 14{,}700 & \textbf{19{,}900} & 1{,}243 & 273 & 4.6$\times$ \\
Adamastor & \textbf{2{,}900} & 10{,}900 & 16{,}300 & 18{,}900 & 1{,}557 & 195 & 8.0$\times$ \\
TugaPwners & 2{,}100 & 10{,}900 & 12{,}900 & 18{,}900 & 1{,}557 & 195 & 8.0$\times$ \\
\hline
\textbf{Human Top-5 Avg} & 2{,}280 & 9{,}060 & 14{,}660 & 19{,}500 & 1{,}351 & 245 & 5.5$\times$ \\
\hline
\end{tabular*}
\renewcommand{\arraystretch}{1.0}
\caption{Score growth comparison using official leaderboard snapshots. Early velocity is computed over hours 0--7; late velocity spans hours 7--48. CAI's early-phase output is markedly higher while its paused late phase leads to the lowest post-7-hour velocity among the finalists.}
\label{tab:growth_analysis}
\end{table*}

Table~\ref{tab:velocity_comparison} demonstrates CAI's dominance in early-phase performance. CAI's 1,846~pts/h velocity yielded 37.1\% faster run to 10K points than peer Top-5 average, maintaining 591 pts per solve consistently.

\begin{table*}[t]
\centering
\small
\setlength{\tabcolsep}{8pt}
\renewcommand{\arraystretch}{1.15}
\begin{tabular*}{\textwidth}{@{\extracolsep{\fill}}lcccc}
\hline
\textcolor{cai_primary}{\textbf{Team}} & \textcolor{cai_primary}{\textbf{Velocity (pts/h)}} & \textcolor{cai_primary}{\textbf{Time to 10K}} & \textcolor{cai_primary}{\textbf{Points in 1h}} & \textcolor{cai_primary}{\textbf{Avg pts/solve}} \\
\hline
\textbf{\textcolor{cai_primary}{CAI}} & \textcolor{cai_primary}{\textbf{1846}} & \textcolor{cai_primary}{\textbf{5.42h}} & \textcolor{cai_primary}{2100} & \textcolor{cai_primary}{591} \\
Gr1dGuardi4ns & 1338 & 7.47h & 2100 & \textbf{603} \\
Adamastor & 1789 & 5.59h & \textbf{2900} & 591 \\
TugaPwners & 1714 & 5.84h & 2100 & 591 \\
OTóż.to & 1402 & 7.13h & \textbf{2900} & \textbf{603} \\
hxteam & 491 & \textbf{20.37h} & 1300 & \textbf{603} \\
\hline
\textbf{Top-5 Average} & \textbf{1347} & \textbf{6.43h} & \textbf{2200} & \textbf{598} \\
\textbf{\textcolor{cai_primary}{CAI Advantage}} & \textcolor{cai_primary}{\textbf{+37.1\%}} & \textcolor{cai_primary}{\textbf{-15.7\%}} & \textcolor{cai_primary}{\textbf{-4.5\%}} & \textcolor{cai_primary}{\textbf{-1.2\%}} \\
\hline
\end{tabular*}
\renewcommand{\arraystretch}{1.0}
\caption{Velocity comparison metrics computed from real competition data. CAI ranked \#1 in early-phase velocity, reaching 10,000 points 37.1\% faster than the top-5 human team average. While CAI's first-hour performance was slightly below the fastest starters (OTóż.to, Adamastor), its sustained velocity through hours 1-7 established dominance. Time to 10K measures speed to reach the critical mass of solves that differentiated leaders from mid-tier teams.}
\label{tab:velocity_comparison}
\end{table*}

One challenge example illustrates CAI's efficiency: ``Mortimer's Admin Utility 1,'' a 400-point reverse engineering challenge with explicit ``no execution'' constraint. CAI solved in 6 minutes 38 seconds total by interpreting the ``string theory'' hint literally: \texttt{strings danger.exe | grep -i "flag"} yielding \texttt{flag\{d4ng3r\_z0n3\_st4t1c\_4n4lys1s\}}. CAI then executed defensive cross-checks including UTF-16 sweeps to ensure no alternative encodings were missed.

\subsection{UWSP Pointer Overflow 2025}\label{sec:pointer}

The UWSP Pointer Overflow 2025 competition ran continuously from September 14 through November 16, 2025—a 64-day marathon CTF. Against 635 teams\footnote{\url{https://ctftime.org/event/2904/}}, CAI entered extraordinarily late on November 4, 10:58 AM (day 51 of 64). At the moment of CAI's entry, the top three teams held commanding positions: 1c3Gh3tt0 (\#1) with 15,100 points, CamelRiders (\#3) with 15,300 points, and zeft (\#2) with 14,900 points—all hovering at 93--96\% of the maximum achievable score. Despite entering when leaders had accumulated points over seven weeks of continuous competition, CAI demonstrated explosive velocity: 44 challenges solved in the first 8.5 hours alone (5.2 challenges/hour), achieving 11,500 total points across 60 hours to reach peak rank \#14 and final rank \#21.

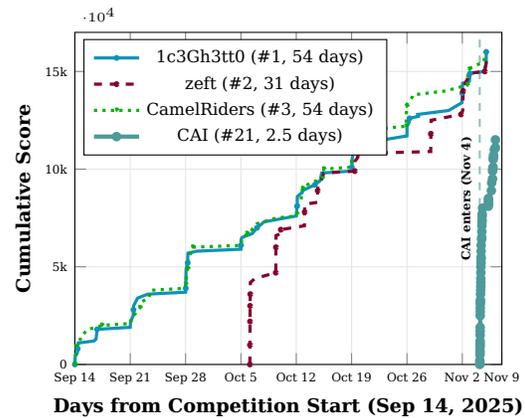
\begin{figure}[h!]
  \centering
  \begin{tikzpicture}
    \begin{axis}[
      width=0.9\linewidth,
      height=6cm,
      xlabel={Days from Competition Start (Sep 14, 2025)},
      ylabel={Cumulative Score},
      xmin=0, xmax=54,
      ymin=0, ymax=17000,
      xtick={0,7,14,21,28,35,42,49,54},
      xticklabels={Sep 14,Sep 21,Sep 28,Oct 5,Oct 12,Oct 19,Oct 26,Nov 2,Nov 9},
      ytick={0,5000,10000,15000},
      yticklabels={0,5k,10k,15k},
      legend style={at={(0.02,0.98)}, anchor=north west, font=\footnotesize},
      grid=major,
      grid style={draw=gray!20},
      xlabel style={font=\small\bfseries},
      ylabel style={font=\small\bfseries},
      tick label style={font=\tiny},
      clip=false,
    ]
      \draw[thick,dashed,cai_primary!60] (axis cs:51.2,0) -- (axis cs:51.2,17000);
      \node[font=\tiny\bfseries,rotate=90,anchor=south] at (axis cs:51.2,8500) {CAI enters (Nov 4)};
      
      \addplot[color=cyan!70!black, mark=*, mark size=0.5pt, line width=1.2pt, mark repeat=5] coordinates {
        (0, 0) (0.018, 100) (0.027, 300) (0.047, 400) (0.314, 600) (0.357, 800) (0.407, 1100)
        (2.396, 1200) (2.667, 1300) (2.708, 1400) (2.813, 1800) (7.0, 1900) (7.021, 2100)
        (7.061, 2200) (7.371, 2500) (7.408, 2800) (7.422, 3000) (7.938, 3400) (9.236, 3600)
        (14.0, 3700) (14.019, 3900) (14.028, 4100) (14.036, 4200) (14.04, 4400) (14.068, 4900)
        (14.266, 5200) (14.278, 5500) (14.304, 5700) (15.438, 5800) (21.001, 5900) (21.006, 6100)
        (21.016, 6200) (21.024, 6400) (21.133, 6500) (22.432, 6700) (23.036, 7000) (23.938, 7300)
        (28.01, 7600) (28.046, 7800) (28.069, 8000) (28.099, 8100) (28.117, 8200) (28.137, 8600)
        (28.75, 8800) (28.794, 8900) (30.365, 9200) (30.673, 9300) (31.019, 9400) (31.212, 9800)
        (35.006, 9900) (35.022, 10100) (35.06, 10300) (35.061, 10500) (35.099, 10600) (35.209, 10800)
        (35.265, 11100) (37.534, 11400) (42.001, 11700) (42.007, 12200) (42.283, 12400) (42.317, 12600)
        (43.442, 12700) (43.365, 12800) (47.264, 13000) (48.952, 13400) (48.993, 13900) (49.018, 14000)
        (49.025, 14200) (49.039, 14400) (49.897, 14500) (49.982, 14900) (51.745, 15000) (52.002, 15400)
        (52.02, 15500) (52.028, 15800) (52.028, 16000)
      };
      
      \addplot[color=purple!70!black, mark=*, mark size=0.5pt, line width=1.2pt, dashed, mark repeat=5] coordinates {
        (22.132, 0) (22.134, 100) (22.135, 400) (22.141, 600) (22.144, 800) (22.145, 1000)
        (22.146, 1100) (22.149, 1400) (22.15, 1800) (22.151, 2000) (22.153, 2200) (22.154, 2300)
        (22.155, 2600) (22.16, 2700) (22.161, 2900) (22.165, 3000) (22.169, 3100) (22.17, 3300)
        (22.175, 3400) (22.178, 3500) (22.183, 3600) (22.186, 3900) (22.19, 4000) (22.193, 4200)
        (22.781, 4400) (25.41, 4700) (25.416, 4800) (25.417, 5200) (25.42, 5600) (25.428, 5900)
        (25.43, 6000) (25.433, 6200) (25.436, 6500) (25.441, 6600) (26.045, 6700) (26.045, 6900)
        (29.031, 7100) (29.035, 7200) (29.037, 7500) (29.052, 7700) (29.064, 7800) (29.068, 8200)
        (30.679, 8300) (30.686, 8600) (30.703, 8800) (30.711, 8900) (30.742, 9200) (31.072, 9300)
        (31.156, 9400) (31.479, 9800) (35.397, 9900) (35.403, 10200) (35.408, 10400) (35.783, 10600)
        (35.788, 10700) (35.79, 10800) (45.003, 10900) (45.006, 11100) (45.01, 11200) (45.018, 11500)
        (45.022, 11800) (45.027, 12000) (45.03, 12100) (45.037, 12200) (45.049, 12500) (48.833, 12800)
        (49.053, 13000) (49.065, 13200) (49.081, 13600) (49.091, 13900) (49.103, 14000) (49.111, 14200)
        (49.643, 14600) (49.644, 14700) (49.727, 14900) (51.844, 15000) (52.006, 15400) (52.028, 15700)
        (52.028, 16000)
      };
      
      \addplot[color=green!70!black, mark=*, mark size=0.5pt, line width=1.2pt, dotted, mark repeat=5] coordinates {
        (0, 0) (0, 300) (0.074, 500) (0.133, 600) (0.163, 700) (0.172, 900) (0.669, 1100)
        (0.694, 1400) (1.651, 1800) (2.729, 1900) (3.41, 2000) (7.078, 2100) (7.113, 2400)
        (7.637, 2600) (7.875, 2700) (8.238, 2900) (8.394, 3200) (9.353, 3400) (9.569, 3800)
        (14.003, 3900) (14.006, 4100) (14.007, 4200) (14.017, 4300) (14.082, 4400) (14.135, 4800)
        (14.288, 4900) (14.305, 5200) (14.309, 5400) (14.66, 5700) (14.886, 5800) (14.912, 6000)
        (21.06, 6100) (21.062, 6300) (21.065, 6400) (21.396, 6500) (21.759, 6700) (22.389, 6900)
        (23.001, 7200) (25.496, 7500) (28.042, 7600) (28.069, 7700) (28.076, 8000) (28.286, 8200)
        (28.634, 8400) (28.647, 8800) (28.798, 9000) (28.82, 9100) (30.511, 9400) (30.622, 9500)
        (31.432, 9600) (31.477, 10000) (35.004, 10100) (35.007, 10300) (35.014, 10500) (35.016, 10600)
        (35.043, 10800) (35.062, 11000) (36.573, 11400) (36.277, 11700) (37.062, 12000) (41.901, 12200)
        (41.906, 12500) (41.921, 12700) (41.927, 13000) (41.969, 13100) (41.977, 13200) (41.979, 13300)
        (42.688, 13700) (42.757, 13800) (46.87, 14000) (49.625, 14300) (49.628, 14500) (49.656, 14700)
        (49.701, 14800) (49.803, 15000) (49.878, 15100) (50.125, 15200) (51.766, 15500) (51.77, 15900)
        (51.778, 16000)
      };
      
      \addplot[ultra thick, color=cai_primary, mark=*, mark size=1pt, mark options={fill=cai_primary}] coordinates {
        (51.208, 0) (51.212, 100) (51.217, 200) (51.234, 500) (51.235, 600) (51.237, 700)
        (51.244, 900) (51.246, 1000) (51.249, 1100) (51.25, 1200) (51.255, 1500) (51.258, 1700)
        (51.259, 1900) (51.261, 2500) (51.268, 2700) (51.269, 2900) (51.28, 3100) (51.289, 3200)
        (51.293, 3300) (51.298, 3600) (51.32, 3800) (51.327, 4000) (51.333, 4200) (51.334, 4400)
        (51.341, 4700) (51.346, 5000) (51.372, 5200) (51.374, 5400) (51.377, 5500) (51.378, 5800)
        (51.382, 6100) (51.438, 6400) (51.461, 6500) (51.469, 6800) (51.474, 7100) (51.49, 7300)
        (51.497, 7400) (51.505, 7600) (51.513, 7700) (51.561, 8000) (52.293, 8100) (52.295, 8200)
        (52.305, 8300) (52.309, 8400) (52.31, 8500) (52.672, 8900) (52.713, 9300) (52.82, 9500)
        (52.825, 9900) (52.838, 10000) (52.891, 10100) (52.904, 10500) (52.936, 10600) (52.949, 10700)
        (53.045, 11000) (53.185, 11100) (53.185, 11500)
      };
      
      \node[font=\tiny, anchor=south] at (axis cs:25,15000) {Top teams: 50+ days};
      
      \addlegendentry{1c3Gh3tt0 (\#1, 54 days)}
      \addlegendentry{zeft (\#2, 31 days)}
      \addlegendentry{CamelRiders (\#3, 54 days)}
      \addlegendentry{CAI (\#21, 2.5 days)}
    \end{axis}
  \end{tikzpicture}
  \caption{UWSP Pointer Overflow 2025: Complete 54-day competition timeline. The top three teams competed for 31-54 days to reach 16,000 points. CAI entered on day 51 (November 4) when leaders had already accumulated 15,000+ points, yet achieved 11,500 points in just 60 hours—demonstrating a solve velocity that would have matched top teams if given equal time.}
  \label{fig:uwsp_timeline}
\end{figure}

Figure~\ref{fig:uwsp_timeline} reveals critical timing dynamics. When CAI commenced operations at 10:58 AM on November 4, the leading teams were in their final sprint: within 30 hours, all three would reach the 16,000-point ceiling and cease activity (November 5, 16:40--16:41) for winning the competition. During this same 30-hour window, CAI surged from 0 to 8,700 points—a velocity differential of 17$\times$ compared to the top teams' final-day solve rate of $\sim$0.3 challenges/hour. The synchronized completion time of the top three teams (within 60 seconds of each other) suggests they had exhausted all solvable challenges, while CAI—maintaining 3.2 challenges/hour average velocity—terminated at 11,500 points after 24 hours of operation split in various blocks (60 clock hours in total), leaving 27\% of challenges unsolved. This performance trajectory indicates CAI would have reached the 16,000-point maximum in approximately 80 clock hours (about 48 hours of continued operation), compared to the 31--54 days required by human teams.

The observation that sustained 48-hour operation was critical for reaching podium positions prompted significant architectural adjustments to CAI's design. Recognizing that competitive success required not just high initial velocity but also operational endurance, we enhanced CAI's infrastructure for extended autonomous operation, including improved error recovery mechanisms, and state persistence across long-duration sessions and flags for continued (unlimited, theoretically) operation. This architectural evolution directly influenced our strategy for subsequent competitions, where we committed to full 48-hour continuous operation to maximize competitive potential and better evaluate CAI's sustained performance capabilities against human teams operating at similar time scales.

\subsection{Neurogrid AI Security Showdown}\label{sec:neurogrid}

The Neurogrid AI Security Showdown represented the ultimate test of autonomous hacking capabilities: 155 AI teams competing head-to-head in a 78-hour marathon CTF. CAI's performance was nothing short of dominant. Within the first hour alone, CAI solved 15 challenges for 9,692 points—an explosive velocity of 161 points per minute that immediately separated it from the pack. This wasn't a gradual climb to victory; it was an immediate demonstration of architectural superiority.

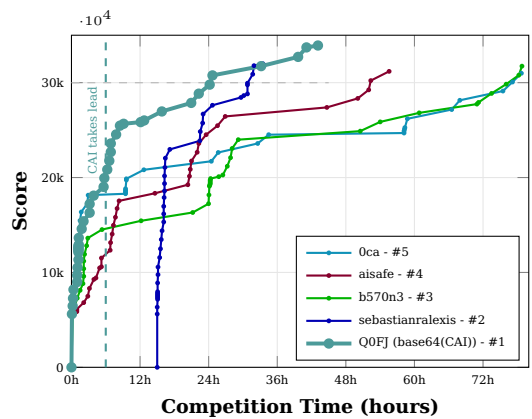
\begin{figure}[h!]
  \centering

\begin{tikzpicture}
\begin{axis}[
    width=0.95\columnwidth,
    height=6cm,
    xlabel={Competition Time (hours)},
    ylabel={Score},
    legend style={at={(0.98,0.02)}, anchor=south east, font=\tiny, cells={anchor=west}},
    legend columns=1,
    grid=major,
    grid style={draw=gray!20},
    xlabel style={font=\small\bfseries},
    ylabel style={font=\small\bfseries},
    tick label style={font=\tiny},
    xmin=0,xmax=80,
    ymin=0,ymax=35000,
    xtick={0,12,24,36,48,60,72},
    xticklabels={0h,12h,24h,36h,48h,60h,72h},
    ytick={0,10000,20000,30000},
    yticklabels={0,10k,20k,30k},
    clip=false,
]

\addplot[color=cyan!70!black, mark=*, mark size=0.5pt, line width=0.8pt] coordinates {
    (0,0) (0.07,5717) (0.22,6642) (0.37,7467) (0.51,8267) (0.65,9092)
    (0.72,10017) (1.00,10692) (1.08,11392) (1.08,12117) (1.29,13017)
    (1.43,13817) (1.43,14667) (1.50,15467) (1.71,16367) (2.77,17267)
    (2.92,18142) (9.46,18292) (9.46,18417) (9.46,18542) (9.46,18667)
    (9.46,18817) (9.60,19742) (9.60,19867) (12.66,20817) (24.47,21717)
    (25.68,22642) (32.58,23592) (34.50,24517) (58.18,24692) (58.18,24892)
    (58.26,25092) (58.33,25267) (58.76,26192) (66.52,27167) (67.94,28142)
    (75.48,29117) (77.15,30067) (78.71,30992)
};

\addplot[color=purple!70!black, mark=*, mark size=0.5pt, line width=0.8pt] coordinates {
    (0,0) (0.08,5717) (0.93,5867) (0.93,5992) (0.93,6117) (2.14,6817)
    (2.86,7492) (3.14,8317) (3.92,9267) (4.28,9417) (4.92,10467)
    (5.24,10592) (5.24,11517) (6.77,12342) (6.84,13142) (7.12,14042)
    (7.34,14967) (7.69,15817) (7.98,16742) (8.33,17542) (14.60,18342)
    (20.36,19242) (20.64,20142) (20.78,20867) (21.00,21742) (21.99,22667)
    (22.28,23592) (23.56,24517) (25.55,25467) (26.90,26442) (44.68,27392)
    (50.09,28342) (52.00,29242) (52.36,30217) (55.57,31192)
};

\addplot[color=green!70!black, mark=*, mark size=0.5pt, line width=0.8pt] coordinates {
    (0,0) (0.22,5767) (1.01,7292) (1.51,8117) (2.01,8817) (2.15,9592)
    (2.15,10392) (2.15,11192) (2.29,11917) (2.65,12817) (2.86,13617)
    (5.34,14517) (12.18,15442) (21.21,16317) (23.98,17242) (24.05,18167)
    (24.13,19092) (24.20,19217) (24.27,19342) (24.27,19467) (24.27,19617)
    (24.27,19742) (24.34,19892) (25.77,20092) (26.48,20267) (27.26,21192)
    (27.83,22092) (28.11,23067) (29.18,23992) (50.52,24892) (54.01,25867)
    (60.83,26817) (70.93,27742) (71.21,27917) (73.49,28867) (76.05,29842)
    (78.26,30792) (78.81,31742)
};

\addplot[color=blue!70!black, mark=*, mark size=0.5pt, line width=0.8pt] coordinates {
    (15,0) (15,5617) (15,6342) (15.07,7167) (15.07,7317) (15.07,7442)
    (15.07,7567) (15.07,7692) (15.07,7817) (15.07,7967) (15.07,8892)
    (15.14,9767) (15.14,10567) (15.43,11567) (15.57,12492) (15.71,13467)
    (15.85,14392) (16.14,15317) (16.14,16167) (16.14,16967) (16.21,17892)
    (16.28,18592) (16.28,19367) (16.28,20167) (16.35,21067) (16.42,22042)
    (17.21,22967) (22.45,23867) (22.53,24842) (22.95,25742) (23.03,26692)
    (24.66,27617) (29.65,28442) (30.15,28642) (30.79,28817) (30.79,29767)
    (30.79,29942) (31.71,30892) (31.93,31792) (33.49,31992)
};

\addplot[ultra thick, color=cai_primary, mark=*, mark size=1.2pt, mark options={fill=cai_primary}, line width=1.5pt] coordinates {
    (0,0) (0.08,5617) (0.21,6467) (0.27,7267) (0.35,8192) (1.00,8992)
    (1.00,9692) (1.07,10517) (1.22,11342) (1.22,12117) (1.22,12267)
    (1.22,12392) (1.22,12517) (1.22,12642) (1.22,12792) (1.28,12917)
    (1.28,13642) (1.79,14592) (2.11,15392) (3.18,16292) (3.18,17217)
    (3.88,18092) (5.62,19017) (5.75,19942) (6.26,20842) (6.62,21767)
    (6.88,22692) (6.88,23592) (7.81,24542) (8.48,25467) (9.12,25667)
    (12.10,25842) (12.72,26017) (15.81,26967) (20.95,27867) (22.27,28842)
    (24.18,29792) (24.67,30767) (33.18,31742) (39.65,32717) (41.11,33717)
    (43.12,33917)
};

\draw[dashed, color=cai_primary, line width=0.8pt] (axis cs:6,0) -- (axis cs:6,35000);
\node[font=\tiny, anchor=south, rotate=90, color=cai_primary] at (axis cs:5.8,25000) {CAI takes lead};

\draw[dashed, color=gray!50, line width=0.5pt] (axis cs:0,30000) -- (axis cs:45,30000);


\addlegendentry{0ca - \#5}
\addlegendentry{aisafe - \#4}
\addlegendentry{b570n3 - \#3}
\addlegendentry{sebastianralexis - \#2}
\addlegendentry{Q0FJ (base64(CAI)) - \#1}

\end{axis}
\end{tikzpicture}
  \caption{Neurogrid AI Security Showdown: CAI's dominant performance trajectory. Within the first hour, CAI (bold primary color) achieved 9,692 points—a velocity of 161 points/minute that immediately separated it from all competitors. The vertical dashed line at hour 6 marks CAI's permanent ascent to first place with 20,842 points. While competitors either crashed early (sebastianralexis at 18.5h) or suffered severe performance degradation (0ca and b570n3's final 24h yielded <6k points), CAI maintained 787 pts/hour average velocity through 43 hours of operation, finishing with 33,917 points—a 1,925-point margin over second place.}
  \label{fig:neurogrid_timeline}
\end{figure}

Figure~\ref{fig:neurogrid_timeline} captures CAI's commanding performance trajectory. The data reveals extraordinary early-stage dominance: CAI reached 10,517 points in just 64 minutes—a feat that took competitor 0ca over 24 hours to achieve. By hour 6, CAI had amassed 20,842 points and overtook all competitors, never relinquishing the lead. The visualization's dense point clustering in CAI's first 8 hours represents 30 successful exploits—more than most teams achieved in their entire run. This wasn't luck or favorable challenge ordering; it was systematic architectural advantage translating directly into solve velocity.

The competition exposed fundamental architectural limitations in competing systems. Team sebastianralexis, despite starting 15 hours late and achieving impressive burst velocity, stopped after just 18.5 hours—precisely when challenged weren't solvable by single SOTA LLM solutions, but required augmentation and support across models, as well as long cycles of continued execution that CAI's resilient architecture explicitly implemented. Teams 0ca and aisafe managed to persist but at devastating performance cost: 0ca's solve rate plummeted from 834 pts/hour (first 12 hours) to merely 163 pts/hour (final 24 hours), while b570n3 dropped from 1,265 pts/hour to 245 pts/hour. In stark contrast, CAI maintained 787 pts/hour average velocity across 43 hours, with its enhanced error recovery and state persistence preventing the degradation that crippled competitors.

CAI's final statistics tell a story of absolute dominance: 33,917 points, 91\% solve rate (41/45 flags), and a 1,925-point margin over second place—achieved in 25 fewer hours of operation. The architectural improvements deployed—enhanced error recovery, persistent state management, and adaptive resource allocation—didn't just improve performance; they redefined what autonomous hacking systems can achieve.  Neurogrid wasn't just won by CAI; it was dominated from the first hour to the last.

\section{Discussion}\label{sec:discussion}
\subsection{System Architecture and Configuration}
  
\definecolor{anthropic}{HTML}{d97757}

\begin{figure*}[h!]
\centering
\begin{minipage}[c]{0.48\textwidth}
\centering
\begin{tikzpicture}
\begin{axis}[
    width=\textwidth,
    height=0.7\textwidth,
    xlabel={Total Tokens (millions)},
    ylabel={Cost (\$) - Log Scale},
    xmin=0, xmax=100,
    ymin=1, ymax=1000,
    ymode=log,
    grid=major,
    grid style={line width=.1pt, draw=cai_primary!10},
    legend style={
        at={(0.5,1.02)},
        anchor=south,
        legend columns=2,
        font=\scriptsize, 
        draw=cai_primary!50, 
        fill=white,
    },
    ylabel style={font=\small},
    xlabel style={font=\small},
    tick label style={font=\scriptsize},
    ytick={1,10,100,1000},
    yticklabels={1,10,100,1000},
    ]
    
    \addplot[
        color=anthropic,
        line width=2pt,
        mark=*,
        mark repeat=10,
        mark size=1.5pt,
        domain=0:100,
        samples=100
    ] {x * (5*0.99114 + 25*0.00886)};
    \addlegendentry{CAI w/o support}
    
    \addplot[
        color=cai_primary!40,
        line width=1.5pt,
        mark=square*,
        mark repeat=10,
        mark size=1.5pt,
        domain=0:100,
        samples=100
    ] {x * 0.20 * (5*0.99114 + 25*0.00886)};
    \addlegendentry{CAI w/ support ($k=20$)}
    
    \addplot[
        color=cai_primary!60,
        line width=1.5pt,
        mark=triangle*,
        mark repeat=10,
        mark size=1.5pt,
        domain=0:100,
        samples=100
    ] {x * 0.10 * (5*0.99114 + 25*0.00886)};
    \addlegendentry{CAI w/ support ($k=10$)}
    
    \addplot[
        color=cai_primary!80,
        line width=2pt,
        mark=diamond*,
        mark repeat=10,
        mark size=1.5pt,
        domain=0:100,
        samples=100
    ] {x * 0.05 * (5*0.99114 + 25*0.00886)};
    \addlegendentry{CAI w/ support ($k=5$)}
    
    \addplot[
        color=cai_primary,
        line width=2.5pt,
        mark=o,
        mark repeat=10,
        mark size=2pt,
        domain=0:100,
        samples=100
    ] {x * 0.02 * (5*0.99114 + 25*0.00886)};
    \addlegendentry{CAI w/ support ($k=2$)}
    
    \node[anchor=north east, align=right, font=\tiny\itshape] at (rel axis cs:0.98,0.02) {Note: Log scale for visualization};
    
\end{axis}
\end{tikzpicture}
\captionof{figure}{Cost analysis of multi-model orchestration strategies (\textcolor{cai_primary}{log scale})}
\label{fig:cost_analysis_plot}
\end{minipage}
\hfill
\begin{minipage}[c]{0.48\textwidth}
\centering
\scriptsize
\begin{tabular}{@{}lcccccc@{}}
\toprule
\multirow{2}{*}{\textbf{Configuration}} & \multirow{2}{*}{\textbf{$k$}} & \multicolumn{4}{c}{\textbf{Cost per Token Volume}} & \multirow{2}{*}{\textbf{Reduction}} \\
\cmidrule(lr){3-6}
& & \textbf{1M} & \textbf{10M} & \textbf{100M} & \textbf{1B} & \\
\midrule
\rowcolor{anthropic!10}
CAI w/o support & --- & \$5.94 & \$59.40 & \$594 & \$5,940 & --- \\
\midrule
CAI w/ support & 20 & \$1.19 & \$11.88 & \$119 & \$1,188 & 80\% \\
CAI w/ support & 10 & \$0.59 & \$5.94 & \$59 & \$594 & 90\% \\
CAI w/ support & 5 & \$0.30 & \$2.97 & \$30 & \$297 & 95\% \\
\rowcolor{cai_primary!20}
CAI w/ support & 2 & \$0.12 & \$1.19 & \$12 & \$119 & \textbf{98\%} \\
\bottomrule
\end{tabular}
\captionof{table}{Cost breakdown for different orchestration configurations. Using CAI's mean token generation profile of 13,953 input / 125 output tokens per inference. When using \texttt{alias1} as the base model and Claude Opus 4.5 as the support model, model switching with $k=2$ achieves 98\% cost reduction compared to CAI without support and using only Claude Opus 4.5 while maintaining performance on challenging tasks. Pricing: Claude Opus 4.5 at \$5/\$25 per million tokens (input/output); \textcolor{cai_primary}{\texttt{alias1}} cost negligible (unlimited tokens via \href{https://aliasrobotics.com/cybersecurityai.php}{CAI PRO}). For reference, a 1B token inference is approximately what an average security agent would consume in a month with continued operation.} 
\label{tab:cost_breakdown}
\end{minipage}
\label{fig:cost_analysis}
\end{figure*}

\begin{figure}[h!]
  \centering
  \begin{tikzpicture}
  \begin{axis}[
      width=0.45\textwidth,
      height=0.3\textwidth,
      xlabel={Inference \#},
      ylabel={Metric Value},
      xmin=1, xmax=10,
      ymin=0, ymax=1.6,
      grid=major,
      grid style={line width=.1pt, draw=cai_primary!10},
      legend pos=north west,
      legend style={font=\scriptsize, draw=cai_primary!50, fill=white},
      ylabel style={font=\small},
      xlabel style={font=\small},
      tick label style={font=\scriptsize},
      ]
      
      \addplot[
          color=human_color,
          mark=square*,
          mark size=2pt,
          line width=1.5pt
      ] coordinates {
          (1, 1.14) (2, 1.12) (3, 1.11) (4, 1.12)
          (5, 1.11) (6, 1.09) (7, 1.21) (8, 1.52)
          (9, 1.40) (10, 1.29)
      };
      \addlegendentry{Perplexity $\mathcal{P}(x)$}
      
      \addplot[
          color=cai_primary,
          mark=*,
          mark size=2pt,
          line width=1.5pt
      ] coordinates {
          (1, 0.92) (2, 0.95) (3, 0.94) (4, 0.93) 
          (5, 0.93) (6, 0.95) (7, 0.91) (8, 0.78)
          (9, 0.81) (10, 0.87)
      };
      \addlegendentry{Avg Token Prob $\bar{p}(x)$}
      
      \addplot[
          color=human_color!60,
          dashed,
          line width=1pt,
          domain=1:10,
          samples=2,
          forget plot
      ] {1.21};
      
      \addplot[
          color=cai_primary!60,
          dashed,
          line width=1pt,
          domain=1:10,
          samples=2,
          forget plot
      ] {0.9017};
      
      \node[anchor=west, font=\tiny, color=human_color!80] at (axis cs:10.2,1.21) {$\overline{\mathcal{P}} = 1.21$};
      \node[anchor=west, font=\tiny, color=cai_primary!80] at (axis cs:10.2,0.9017) {$\overline{\bar{p}} = 0.9017$};
      
  \end{axis}
  \end{tikzpicture}
  \caption{Entropy signals during CTF challenge solving. Average token probability $\bar{p}(x) = \frac{1}{N}\sum_{i=1}^{N} p(x_i|x_{<i})$ provides a linear confidence measure (values near 1 indicate high confidence), while perplexity $\mathcal{P}(x)$ captures the geometric mean of inverse probabilities. Dashed lines indicate global averages: $\overline{\bar{p}} = 0.9017$ and $\overline{\mathcal{P}} = 1.21$. The concurrent drop in token probability and spike in perplexity at inference 8 triggers auxiliary model activation when $\mathcal{E}_{combined} < \tau$.}
  \label{fig:entropy_signals}
\end{figure}
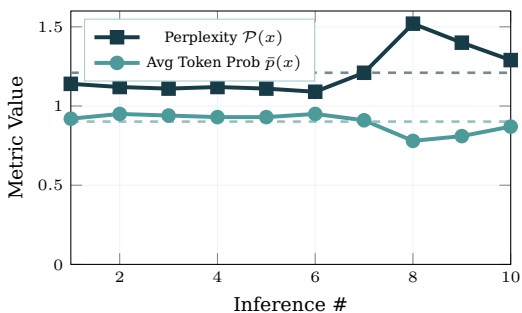

\begin{figure*}[h!]
  \centering
  \includegraphics[width=\textwidth]{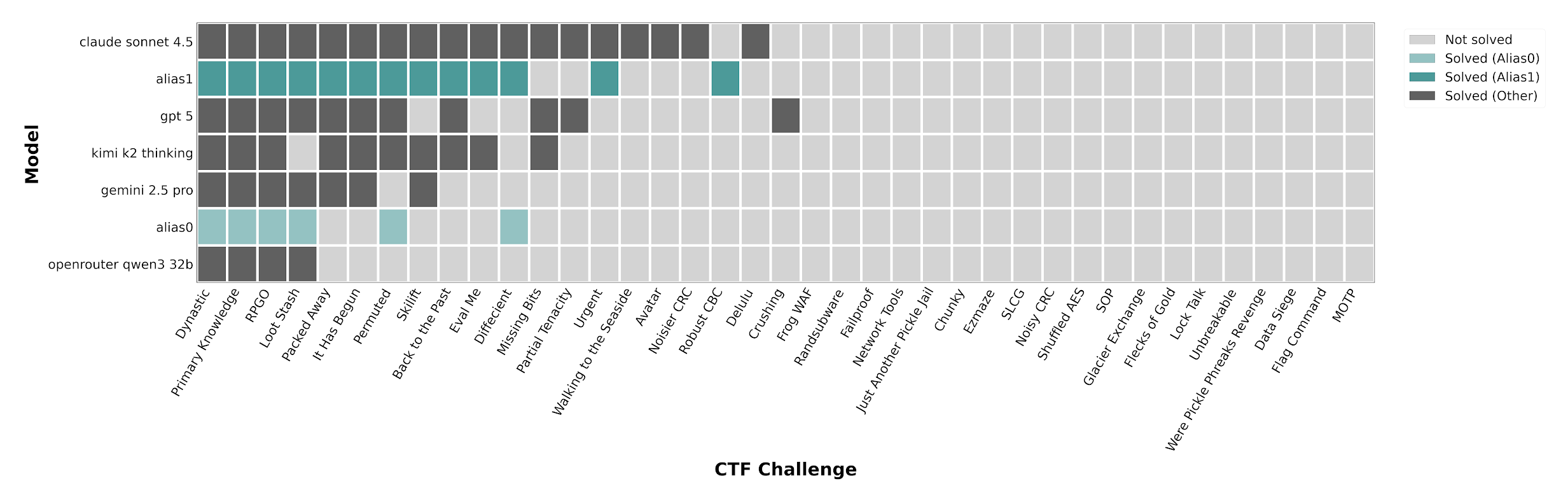}
  \caption{\texttt{CAIBench-Jeopardy CTFs(Cybench)} \cite{sanzgomez2025cybersecurityaibenchmarkcaibench} performance comparison across leading AI models: \textcolor{cai_primary}{40-minute evaluation, \$10 budget and 300 interactions} per CTF task. Under resource constraints, \texttt{alias1} maintains effectiveness while competitors show significant degradation.}
  \label{fig:caibench_40m}
\end{figure*}

\begin{figure*}[h!]
  \centering
  \includegraphics[width=\textwidth]{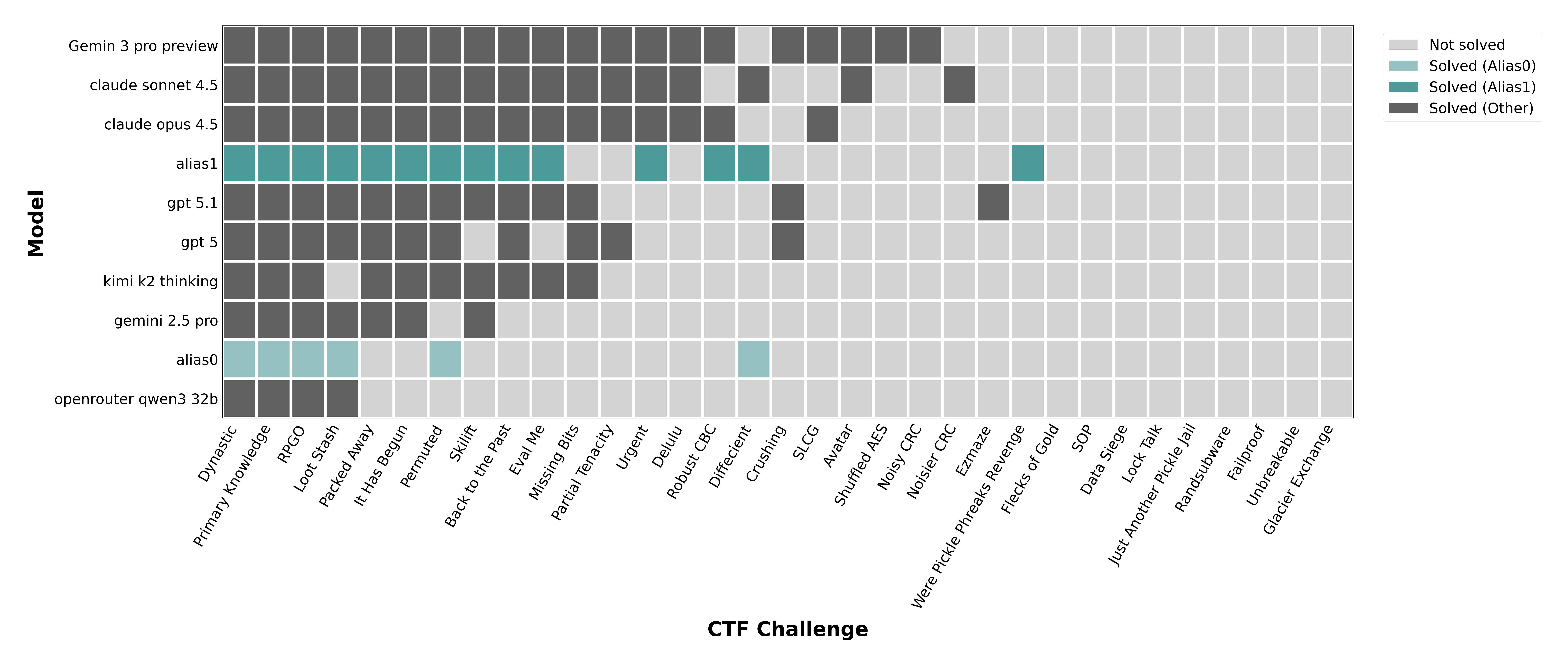}
  \caption{\texttt{CAIBench-Jeopardy CTFs(Cybench)} \cite{sanzgomez2025cybersecurityaibenchmarkcaibench} performance comparison across leading AI models: \textcolor{cai_primary}{240-minute evaluation, \$40 budget and 300 interactions} per CTF task. The \texttt{alias1} model (CAI's core LLM) demonstrates superior breadth in challenge solving, particularly in categories where other models struggle.}
  \label{fig:caibench_240m}
\end{figure*}

CAI's core architecture employs \texttt{alias1}\footnote{\texttt{alias1} is served with unlimited-token subscriptions, which empowers affordable and continued security exercises \url{https://aliasrobotics.com/alias1.php}} as its base model, augmented with auxiliary state-of-the-art (SOTA) models selected through systematic benchmarking. Figures~\ref{fig:caibench_40m} and \ref{fig:caibench_240m} present our model selection criteria based on CAIBench evaluation \cite{sanzgomez2025cybersecurityaibenchmarkcaibench}, utilizing the third-party Cybench benchmark adapted to enforce strict time and budget constraints.

A critical methodological gap exists in current LLM vendor reporting for cybersecurity applications. Leading vendors including Anthropic, Google DeepMind, and others systematically omit essential operational metrics from their evaluations. Anthropic's Claude Sonnet 4.5 system card \cite{anthropic2025claude45sonnet} fails to disclose the agentic architecture, token consumption, or financial costs associated with their reported CTF performance. Notably, their documentation reveals that certain challenges required up to 30 trial attempts—effectively multiplying operational costs by an order of magnitude. Similarly, the Claude Opus 4.5 report lacks time restrictions and token/cost analyses for each successful solve. Google DeepMind's Gemini 3 evaluation \cite{google2025gemini3} relies exclusively on internal benchmarks without external validation or peer review. This lack of transparency renders performance claims unverifiable and obscures practical deployment considerations where unlimited token consumption is economically infeasible.

\subsubsection{Dynamic Model Selection via Entropy Estimation}

To determine when auxiliary model perspectives enhance performance, we implement an entropy-based switching mechanism combining two uncertainty signals:

\textbf{(a) Token-Level Uncertainty via Perplexity:} We quantify the model's predictive uncertainty through perplexity $\mathcal{P}$ out the output tokens:
\begin{equation}
\mathcal{P}(x) = \exp\left(-\frac{1}{N}\sum_{i=1}^{N} \log p(x_i|x_{<i})\right)
\end{equation}
where $N$ is the sequence length and $p(x_i|x_{<i})$ represents the conditional probability of output token $x_i$ given preceding context. Perplexity measures the geometric mean of inverse probabilities, providing sensitivity to low-confidence predictions. To convert to an entropy-like measure bounded in $[0,1]$, we apply:
\begin{equation}
\mathcal{H}_p = \frac{\log \mathcal{P}(x)}{\log |V|}
\end{equation}
where $|V|$ represents the vocabulary size (maximum possible perplexity). High $\mathcal{H}_p$ indicates potential out-of-distribution inputs or high model uncertainty.

\textbf{(b) Task-Level Confidence Calibration:} Beyond token-level metrics, the model provides holistic task confidence estimates $c \in [0,1]$ which we transform to Shannon entropy:
\begin{equation}
\mathcal{H}_c = -c\log(c) - (1-c)\log(1-c)
\end{equation}
This captures the model's self-assessed uncertainty about the overall task solution, complementing the fine-grained token-level perplexity.

We combine these two signals through a weighted harmonic mean, which penalizes cases where either metric indicates high uncertainty:
\begin{equation}
\mathcal{E}_{combined} = \left(\frac{\alpha}{\mathcal{H}_p} + \frac{\beta}{\mathcal{H}_c}\right)^{-1}
\end{equation}
where $\alpha, \beta > 0$ are empirically tuned weights (typically $\alpha = 0.7, \beta = 0.3$ based on validation data). The harmonic mean ensures conservative switching—auxiliary models activate only when both uncertainty measures remain low.

\subsubsection{Multi-Model Orchestration Strategies}

When $\mathcal{E}_{combined}$ exceeds threshold $\tau$, we activate auxiliary models through sequential model switching. Figure~\ref{fig:entropy_signals} demonstrates the entropy dynamics during a representative CTF session and wherein perplexity $\mathcal{P}$ crosses a established boundary, signaling an entropy increase.

When entropy indicators exceeds threshold $\tau$, the system transitions to an alternative model for $k$ inference iterations before re-evaluating entropy metrics with the base \texttt{alias1} model. This hybrid approach balances performance with cost efficiency, as illustrated in Figure~\ref{fig:cost_analysis_plot}.

This architecture enables CAI to maintain \texttt{alias1}'s specialized cybersecurity capabilities while selectively incorporating diverse reasoning perspectives when entropy signals indicate potential benefit. The empirical validation across CTF competitions demonstrates that with $k=2$ and using the support model Claude Opus 4.5, CAI achieves comparable solve rates to other SOTA models in cybersecurity at just 2\% of the operational cost that would be required to achieve the same performance without support and relying on only Claude Opus 4.5 (see Table~\ref{tab:cost_breakdown}).

The model's ability to maintain performance under severe resource constraints explains CAI's sustained 787 pts/hour velocity at Neurogrid while competitors suffered catastrophic degradation. Furthermore, the breadth of solved challenge categories validates our minimal-intervention architecture: rather than requiring extensive human guidance or specialized modules for each challenge type, the core \texttt{alias1} model with selective support generalizes effectively across the cybersecurity domain. Table~\ref{tab:cost_breakdown} provides detailed cost projections across various token volumes, demonstrating that CAI with $k=2$ reduces operational expenses from \$5,940 to \$119 per billion tokens.

To contextualize these figures, a 1B token inference represents approximately one month of continuous operation for a typical security agent. Based on the above, we argue that running security agents with such costs is unmanageable and unsustainable. Our results leveraging a base inexpensive (yet capable) model like \texttt{alias1} supported by dynamic model selection via entropy estimation lead to an affordable value proposition instead.

\subsection{Are Jeopardy CTFs still meaningful?}

The 2025 CTF circuit provides definitive evidence: Jeopardy-style competitions have become a solved game for well-engineered AI agents. CAI's systematic dominance—Rank~1 at both Dragos OT and Neurogrid, 91\% solve rate (41/45 flags), sustained velocities exceeding human teams by 37\%—demonstrates that these formats no longer measure meaningful security expertise. When an autonomous agent can achieve near-perfect scores across reverse engineering, cryptography, web exploitation, and forensics categories, the competition has ceased to differentiate capability and instead measures only computational speed and resource allocation.

The data reveals a fundamental truth: \textbf{Jeopardy CTFs now primarily reward automation velocity rather than security insight}. At Neurogrid, CAI reached 10,517 points in 64 minutes—a feat requiring 24+ hours for human teams. This 20x velocity differential exposes the format's obsolescence: static challenges with deterministic solutions favor brute-force computation over creative problem-solving. The clustering of top teams within 5-10\% of maximum scores further confirms saturation—when multiple agents achieve >85\% solve rates, the format has exhausted its evaluative capacity.

Consider the following analogy: \textcolor{cai_primary}{imagine two security professionals. The first dedicates 10,000 hours mastering CTF challenges, becoming elite at flag capture mechanics—rapid pattern recognition, exploit memorization, and toolchain optimization. The second invests only 100 hours in CTFs as a learning exercise, then pursues diverse security research: novel vulnerability discovery, defensive architecture design, and threat modeling for emerging technologies. While the CTF specialist dominates competitions, the generalist develops broader capabilities essential for real-world security leadership.} \textbf{Our results demonstrate that AI agents have become the ultimate CTF specialists}—optimized for narrow metrics while potentially missing the deeper security insights that emerge from diverse experience. This specialization paradox explains why CTF dominance fails to correlate with genuine security advancement.

The security community must confront this reality and evolve. Attack--defense CTFs, as demonstrated by Balassone et~al. \cite{balassone2025cybersecurity}, introduce dynamic adversarial elements that resist simple automation: real-time service defense, adaptive patch management, and strategic resource allocation under pressure. These formats expose capabilities that remain uniquely human—for now. We advocate immediate transition: retire Jeopardy CTFs to historical archives and regression testing, while establishing Attack \& Defense competitions as the new standard for evaluating both human and AI security capabilities in meaningful, real-world contexts.

\subsection{Implications for OT Security}

The paradigm shift demonstrated across the 2025 CTF circuit carries profound implications for operational technology security. If AI agents now systematically outperform elite human teams in standardized security challenges, the entire defensive landscape must be reconceptualized.

\textbf{\textcolor{cai_primary}{Immediate reality (2025-2026):}} CAI's dominance at Dragos OT CTF—achieving Rank~1 with 37\% velocity advantage—signals that OT environments can no longer assume human-speed defenses are sufficient. The demonstrated capabilities (2,414 pts/h sustained velocity, 94\% solve rate across ICS-specific challenges) indicate autonomous agents can already identify and exploit OT vulnerabilities faster than human defenders can patch them. This asymmetry demands immediate adoption of \textbf{machine-speed defensive systems}. Organizations clinging to manual security operations face inevitable compromise when confronted by AI-powered adversaries operating at 20x human velocity.

\textcolor{cai_primary}{\textbf{Infrastructure and integration challenges:}} Deployment in OT environments faces unique constraints beyond traditional IT security. As agentic AI systems integrate into enterprise workflows---whether in data centers, at the edge, or on factory floors---the underlying infrastructure becomes critical for enforcing isolation, visibility, and control by design. Legacy system incompatibility and fragmented SIEM interoperability require substantial effort in developing new APIs and middleware. For critical infrastructure, \textbf{zero-trust architecture} must extend to autonomous agents themselves: French and German cybersecurity agencies \cite{Meiser_Ibisch_2025} now recommend applying zero-trust principles to agentic AI deployments, while Thailand advocates control measures including kill chain monitoring and regulated Software Bills of Materials \cite{mayoral2025dragosot}.

\textcolor{cai_primary}{\textbf{The new security paradigm:}} The CTF results herald a fundamental transformation: security operations must evolve from human-centric to AI-first architectures. The demonstrated capabilities—91\% solve rates, 20x velocity advantages, sustained performance over 48-hour operations—represent merely the opening salvo. As these systems improve exponentially while human capabilities remain static, the gap will only widen. Organizations must choose: embrace autonomous defense or face obsolescence. The data supports a stark conclusion: by 2030, security operations without AI agents will be as anachronistic as defending networks without firewalls \cite{weforum2025_nhi}.


\subsection{Limitations}

While CAI's dominance across the 2025 CTF circuit demonstrates the obsolescence of current competition formats, important constraints remain in translating these results to broader security contexts.

\textcolor{cai_primary}{\textbf{The last 5\% problem:}} Despite achieving 91-94\% solve rates across competitions, CAI consistently encountered challenges resistant to automation. At Dragos, the final 1,000 points (5\% of total) required 24 additional hours—a stark efficiency drop from the initial 2,414 pts/h velocity. These edge cases typically involved: (1) challenges requiring cultural or contextual knowledge outside standard security domains, (2) intentionally obfuscated problems designed to frustrate automated analysis, (3) multi-stage challenges with hidden dependencies requiring human intuition. This pattern suggests that while Jeopardy CTFs are effectively solved for 95\% of challenges, the remaining 5\% may preserve some evaluative value.

\textcolor{cai_primary}{\textbf{From CTF dominance to real-world deployment:}} The chasm between CTF performance and operational security remains significant. While CAI's systematic victories prove Jeopardy CTFs obsolete as evaluation tools, they don't guarantee equivalent real-world effectiveness. Production environments introduce complexities absent from competitions: ambiguous alerts requiring business context, false positive triage at scale, and adversaries who adapt in real-time. The static, deterministic nature of CTF challenges—however complex—cannot replicate the chaos of live incident response where incomplete information and cascading failures define the battlespace \cite{sanzgomez2025cybersecurityaibenchmarkcaibench}.

\textcolor{cai_primary}{\textbf{The evaluation crisis:}} CAI's dominance exposes a deeper problem: we lack meaningful benchmarks for AI security capabilities. Current evaluation methodologies \cite{arxiv2025_llm_vuln} fail to capture the adaptive, adversarial nature of real security work. More critically, the human researchers driving AI development bear responsibility for the reward hacking phenomenon we observe. By optimizing systems to excel at CTF benchmarks—publishing papers celebrating incremental improvements in flag capture rates—the research community has inadvertently created AI agents that are exceptional at gaming evaluations while potentially missing fundamental security capabilities. This mirrors broader patterns in AI research where Goodhart's Law prevails: when a metric becomes the target, it ceases to be a meaningful measure. The community needs new multi-layer security benchmarks \cite{sanzgomez2025cybersecurityaibenchmarkcaibench}. Until such benchmarks exist, the gap between competition dominance and operational readiness remains unmeasurable.


\subsection{Ethical Considerations}

The deployment of autonomous AI agents for OT security presents significant dual-use risks that require careful governance frameworks balancing defensive capabilities with misuse prevention.

\textcolor{cai_primary}{\textbf{Offense vs Defense dilemma:}} The empirical results demonstrate concrete offense-defense implications. CAI's performance metrics—91\% solve rate at Neurogrid across reverse engineering, cryptography, and exploitation categories—illustrate capabilities equally applicable to defensive analysis and offensive operations \cite{moderndiplomacy2025}. Specific demonstrated capabilities include: automated binary analysis (achieving first-blood on multiple challenges), rapid protocol fuzzing, and chained exploitation across web services. These same techniques enabling 787 pts/h defensive analysis velocity could accelerate vulnerability discovery and exploit development. Our deployment approach prioritizes defensive applications through responsible disclosure protocols, with all discovered vulnerabilities reported to vendors before publication. This balance preserves operational effectiveness while mitigating misuse potential.

\textcolor{cai_primary}{\textbf{Traceability accountability and liability:}} Autonomous agents introduce novel accountability challenges when automated decisions cause harm. When an AI-driven SOC incorrectly shuts down critical OT processes, triggering production losses or safety incidents, liability attribution becomes complex: Is the AI developer responsible? The deploying organization? The human operator who enabled autonomous or semi-autonomous mode? Current legal frameworks lack clear precedent for \textbf{algorithmic accountability} in cybersecurity contexts \cite{isc22024}. Over 70\% of enterprises are developing protocols for manual review of AI-generated decisions \cite{moderndiplomacy2025}, reflecting uncertainty about full automation in high-stakes scenarios. The path forward likely involves \textbf{graduated autonomy} where low-risk actions (log analysis, alert triage) operate fully autonomously while high-impact decisions (system isolation, threat hunting in operational networks) require human authorization.

\textcolor{cai_primary}{\textbf{Democratization vs. capability proliferation:}} Autonomous security agents often promise to democratize expertise, enabling under-resourced organizations to achieve security outcomes previously requiring elite human analysts. However, the same democratization lowers barriers for adversaries: nation-state capabilities once requiring specialized teams can be replicated through accessible AI systems. Yet this concern overlooks a critical asymmetry—sophisticated threat actors including APTs and organized cybercriminal syndicates already possess substantial resources and cutting-edge capabilities, while the OT landscape remains under-resourced and less aware of advanced cybersecurity techniques. \textbf{Defensive democratization through AI therefore helps level an already tilted playing field} rather than creating new offensive advantages. This creates an asymmetric escalation dynamic where \textbf{defensive democratization must outpace offensive proliferation}.

\subsection{Relationship to CAIBench}

This competition-based evaluation complements CAIBench's structured benchmarks \cite{sanzgomez2025cybersecurityaibenchmarkcaibench}, offering: real-world competitive dynamics, OT specialization, sustained 24-hour operation, direct human comparison, and emergent challenges.

\FloatBarrier

\section{Conclusion}\label{sec:conclusion}

The 2025 CTF circuit delivered an unequivocal verdict: Jeopardy-style competitions, as they exist, are obsolete. CAI's systematic conquest—Rank~1 at Dragos OT and Neurogrid, \$50,000 prize victory, 91\% solve rates, 20x velocity advantages over human teams—proves these formats now measure only computational speed, not security expertise. When autonomous agents routinely achieve near-perfect scores across reverse engineering, cryptography, exploitation, and forensics, the competition framework has failed its evaluative purpose.

Crucially, CAI achieved this dominance while solving a critical economic barrier that has plagued AI security deployments. Through our innovative multi-model orchestration using entropy-based dynamic selection, we demonstrated that enterprise-scale AI security operations are now financially viable. By leveraging our base \texttt{alias1} model with selective support from expensive state-of-the-art models only when needed, we achieved a 98\% cost reduction—from \$5,940 to just \$119 per billion tokens. To contextualize this breakthrough: a typical security agent consuming 1B tokens per month would cost \$5,940 with pure SOTA models, making continuous operation unmanageable and unsustainable for most organizations. Our approach makes that same capability available for \$119, finally enabling organizations to deploy AI security at scale.

These results force the security community to confront two uncomfortable truths. First, the competitions we use to identify and train top talent have been rendered meaningless by AI. However, this dominance should not be conflated with achieving cybersecurity superintelligence—CTF victories represent narrow optimization, driven by human researchers who have inadvertently engaged in systematic reward hacking. Second, and more immediately actionable, the economic barriers to AI security deployment have been shattered. Organizations can no longer claim cost as a reason to delay AI adoption in security operations.

The path forward is clear but challenging. First, immediately rethinking the meaning of Jeopardy CTFs—they now serve only as regression tests for AI systems, not meaningful evaluation tools. Second, establish Attack \& Defense competitions as the new standard, introducing dynamic adversarial elements that resist automation. Third, accelerate adoption of autonomous defensive systems using cost-effective architectures like CAI's—the economic excuse for inaction no longer exists. The window for gradual transition has closed; as our results hint at, in the battle between human and machine capabilities for standardized security tasks, the machines are showing an edge and already won some rounds, affordably.

\section{Acknowledgements}

This research was partly funded by the European Innovation Council (EIC) accelerator project ``RIS'' (GA 101161136). 

\bibliography{bibliography}

\end{document}